\documentclass[10pt,journal,doublecolumn]{IEEEtran}
\usepackage{amsmath,amssymb,threeparttable,placeins,enumerate,caption}
\usepackage{mathtools,relsize}

\usepackage{graphicx}
\usepackage{epstopdf}
\usepackage[usenames]{xcolor}
\usepackage{amsfonts}
\usepackage{latexsym}
\usepackage{subfigure,multirow,makecell,stfloats}

\usepackage{algorithm}
\usepackage[noend]{algpseudocode}

\newtheorem{theorem}{{{\textit{Theorem}}}}

\newtheorem{lemma}{{{\textit{Lemma}}}}
\newtheorem{corollary}{{{{\textit{Corollary}}}}}

\newtheorem{definition}{{{\textit{Definition}}}}

\newtheorem{remark}{{{\textit{Remark}}}}

\newtheorem{example}{{{\textit{Example}}}}

\newtheorem{construction}{{{\textit{Construction}}}}
\begin{document}
	
	
	\title{New Sets of Optimal Odd-length Binary Z-Complementary Pairs}
	\author{Avik~Ranjan~Adhikary,~
		Sudhan Majhi,
		Zilong~Liu,~
		Yong~Liang~Guan~
		\thanks{A. R. Adhikary was with the Department of Mathematics, Indian Institute of Technology Patna, India. He is now with Department of Mathematics, Southwest Jiaotong University, China, e-mail: {\tt avik.adhikary@ieee.org}. S. Majhi is with the Department of Electrical Engineering, Indian Institute of Technology Patna, India, e-mail: {\tt smajhi@iitp.ac.in}. Z. Liu was with the School of Electrical and Electronics Engineering, Nanyang Technological University (NTU), Singapore; He is now with the 5G Innovation Centre, Institute for Communication Systems, University of Surrey, UK, e-mail: {\tt zilong.liu@surrey.ac.uk}. Y. L. Guan is with the School of EEE, NTU, Singapore, e-mail: {\tt eylguan@ntu.edu.sg}.}}

\maketitle
	
	\begin{abstract}
		A pair of sequences is called a Z-complementary pair (ZCP) if it has zero aperiodic autocorrelation sums (AACSs) for time-shifts within a certain region, called zero correlation zone (ZCZ). \textit{Optimal} odd-length binary ZCPs (OB-ZCPs) display closest correlation properties to Golay complementary pairs (GCPs) in that each OB-ZCP achieves maximum ZCZ of width $(N+1)/2$ (where $N$ is the sequence length) and every out-of-zone AACSs reaches the minimum magnitude value, i.e. $2$.
		Till date, systematic constructions of \textit{optimal} OB-ZCPs exist only for lengths $2^{\alpha} \pm 1$, where $\alpha$ is a positive integer. In this paper, we construct \textit{optimal} OB-ZCPs of \textit{generic} lengths  $2^\alpha 10^\beta 26^\gamma +1$ (where $\alpha,~ \beta, ~ \gamma$ are non-negative integers and $\alpha \geq 1$) from inserted versions of binary GCPs. The \textit{key} leading to the proposed constructions is several newly identified structure properties of binary GCPs obtained from Turyn's method. This \textit{key} also allows us to construct OB-ZCPs with possible ZCZ widths of $4 \times 10^{\beta-1} +1$, $12 \times 26^{\gamma -1}+1$ and $12 \times 10^\beta 26^{\gamma -1}+1$ through proper insertions of GCPs of lengths $10^\beta,~ 26^\gamma, \text{and } 10^\beta 26^\gamma$, respectively. Our proposed OB-ZCPs have applications in communications and radar (as an alternative to GCPs).
		
	\end{abstract}
	
	\section{Introduction}
	\subsection{Background}
	\IEEEPARstart{T}{he}
	concept of ``complementary pair", also known as Golay complementary pair (GCP), was introduced by Marcel J. E. Golay in the early 1950s. His objective was to design an infrared multislit spectrometry to allow desired radiation with a fixed single wavelength passing through background radiation with many different wavelengths \cite{golay1}. Formally, a pair of sequences is called a GCP if their aperiodic auto-correlations sums (AACSs) are zero for all non-zero time-shifts \cite{golay1,golay2}. After Golay's finding, intensive research activities have been carried out concerning the structures, the constructions, and the applications of GCPs \cite{survey2016}. Some important application of the GCPs include: peak-to-mean envelope power ratio (PMEPR) reduction of multicarrier signals \cite{popovic}, \cite{davis}, Doppler resilient radar waveforms \cite{doppler_gcp}, channel estimation in inter-symbol interference (ISI) channels \cite{spasojevic}, intercell interference rejection \cite{WHMow}, etc. %
	It is noted that binary GCPs are only possible for certain even sequence lengths. The existing known binary GCPs have lengths of the form $2^\alpha 10^\beta 26^\gamma $ only, where $\alpha, ~ \beta, ~ \gamma$ are non-negative integers. To be specific, all the admissible lengths up to 100 for binary GCPs \cite{Borwein03} are
	\begin{displaymath}
	2,4,8,10,16,20,26,32,40,52,64,80.
	\end{displaymath}
	
	To find an alternative of GCPs, in 2007, Fan \textit{et. al.} proposed aperiodic Z-complementary pairs (ZCPs) \cite{fan}, which may be used in the scenarios where the required sequence lengths are not in the form of $2^\alpha 10^\beta 26^\gamma $. An aperiodic ZCP has zero AACSs for certain out-of-phase time-shifts around the in-phase position, called zero correlation zone (ZCZ) \cite{fan}. 
	Based on some computer search results, a conjecture was left in \cite{fan} that for any binary ZCP of odd-length $N$, it has maximum ZCZ width of $(N+1)/2$. In 2011, Li \textit{et al.} \cite{li2010} proved that this conjecture is true, but they did not succeed in finding a systematic construction for odd-length binary ZCPs with ZCZ width $(N+1)/2$.
	
	In 2014, Liu \textit{et al.}  developed a systematic construction of \textit{optimal} aperiodic odd-length binary ZCPs (OB-ZCPs) \cite{zilong_obzcp} by applying insertion method on GCPs with lengths $2^\alpha$ ($\alpha$ non-negative). Each of these pairs is \textit{optimal} in two aspects: it has maximum ZCZ width, i.e., $(N+1)/2$, and every AACS outside the ZCZ has the minimum possible magnitude of 2. Hence, optimal OB-ZCPs display closest correlation properties to GCPs and thus they may be the best sequence pairs to play the role of GCPs whenever odd sequence lengths are required. Two classes of OB-ZCPs have been studied in \cite{zilong_obzcp}: Type-I OB-ZCP, a conventional ZCP having ZCZ for time-shifts around the in-phase position and Type-II OB-ZCP having ZCZ for time-shifts around the end-shift position (away from the in-phase position). Type-I OB-ZCPs can be applied in quasi-synchronous CDMA (QS-CDMA) systems for the mitigation of ISI and multiuser interference \cite{WHMow}, \cite{fan_1}, \cite{fan_2}, while Type-II OB-ZCPs may have potential application for interference rejection in broadband wireless communication systems when the minimum interfering-signal delays are very large \cite{lee2010mobile}. {Type-I and Type-II OB-ZCPs have also been used in designing new sets of complementary sequences \cite{Avik-cl2019} and Z-complementary sequence sets \cite{Avik-el2019}.} From the perspective of combinatorial design, \cite{zilong_obzcp} also points out that each optimal OB-ZCP corresponds to a base-two almost difference families (ADF) \cite{Ding20084941,Ding08-DF}. The periodic analogy of \textit{optimal} OB-ZCPs are the \textit{optimal} binary periodic almost-complementary pairs (BP-ACPs) which have been investigated in \cite{Avik}. BP-ACPs can be employed as \textit{optimal} training sequences for single-carrier multi-antenna frequency-selective fading communication systems \cite{yuan_tu_fan}. Recently, Chen proposed a direct construction of binary and non-binary ZCPs in \cite{chen_novel} based on generalized Boolean functions. It is worthy to note that Chen's ZCPs have lengths in the form of $2^\alpha+2^\beta$ and include certain optimal OB-ZCPs in \cite{zilong_obzcp} and even-length ZCPs (EB-ZCPs) \cite{Liu-SPL2014} as special cases\footnote{Some recent advances on Z-complementary pairs/sets can be found in \cite{Avik-SPL2018,wu-SPL2018,Xie-SPL2018}.}.
	
	\subsection{Contributions}
	Motivated by \textit{optimal} OB-ZCPs of length $2^\alpha +1$ obtained from inserted versions of binary GCPs of length $2^\alpha$, we aim at systematic constructions of \textit{optimal} OB-ZCPs with generic lengths of $2^\alpha 10^\beta 26^\gamma +1$. Unlike the binary GCPs used in \cite{zilong_obzcp} which are constructed from generalized Boolean functions (and thus have lengths of $2^\alpha$ only) proposed by Davis and Jedwab \cite{davis}, we use Turyn's method \cite{Turyn74} to generate binary GCPs of lengths $2^\alpha 10^\beta 26^\gamma$. Several \textit{intrinsic} structure properties of the GCPs from Turyn's method are identified, for the first time. Specifically, these structure properties enable us to identify all the columns of binary GCPs (from Turyn's method) each having identical sign (and opposite signs), when they are arranged as two-dimensional matrices of order $2\times N$. Shown in \textit{Theorem 1} and \textit{Corollaries 1-4}, these structure properties play a \textit{key} role in proving the proposed \textit{optimal} OB-ZCPs of lengths $2^\alpha 10^\beta 26^\gamma+1$ which are obtained by applying insertion method to binary GCPs of lengths $2^\alpha 10^\beta 26^\gamma$, where $\alpha\geq1$. It is noted that these new \textit{optimal} OB-ZCPs (with $\beta+\gamma\geq1,\alpha\geq 1$) cannot be obtained by the approaches in \cite{zilong_obzcp} and some optimal OB-ZCPs in \cite{zilong_obzcp} (with $\beta=\gamma=0$) may be viewed as a special case of our proposed ones. When insertion is applied to binary GCPs of lengths $10^\beta,~ 26^\gamma, \text{ and } 10^\beta 26^\gamma$, we show that the \textit{largest} possible ZCZ widths are $4 \times 10^{\beta -1}+1$, $12 \times 26^{\gamma -1}+1$ and $12 \times 26^{\gamma -1}10^\beta +1$, respectively. {The result in \textit{Corollary 2} has been presented in \cite{Avik-iwsda2017}}.
	
	\subsection{Organization}
	This paper is organized as follows. In Section II, we introduce Type-I and Type-II binary ZCPs, \textit{optimal} OB-ZCPs and the insertion method on sequences. In Section III, we recall the direct form of Turyn's method for binary GCPs and reveal some intrinsic structure properties of these pairs by some element-wise calculations. We present systematic approaches to construct new sets of \textit{optimal} OB-ZCPs by applying insertion method to GCPs of length $2^\alpha 10^\beta 26^\gamma$, $\alpha\geq1,~\beta,\gamma\geq0$ are integers. We show that when insertion method is employed on GCPs of length $10^\beta$, $26^\gamma$ and $10^\beta26^\gamma$, we can achieve a maximum ZCZ width of $4 \times 10^{\beta-1} +1$, $12 \times 26^{\gamma -1}+1$, and $12 \times 10^\beta 26^{\gamma -1}+1$, respectively. We conclude the paper in Section IV.
	
	\section{Preliminaries}
	Throughout this paper, a binary sequence is a vector over alphabet set $\mathbb{U}=\{+1,-1\}$.  ``$\mathbf{a} || \mathbf{b}$" denotes the horizontal concatenation of row sequences $\mathbf{a}$ and $\mathbf{b}$. $\overleftarrow{\mathbf{a}}$ is the reverse of $\mathbf{a}$. $+$ and $-$ denote $1$ and $-1$, respectively. Denote by $\mathbf{c}_L$ a length-$L$ vector with identical entries of $c$. $\mathbf{a} \otimes \mathbf{b}$ denotes the Kronecker product of the sequences $\mathbf{a}$ and $\mathbf{b}$. For sequences $\mathbf{a}$ and $\mathbf{b}$ of identical length $N$, we denote by $(\mathbf{a};\mathbf{b})$ as a $2\times N$ matrix whose two rows are $\mathbf{a}$ and $\mathbf{b}$, respectively. For two length-$N$ binary sequences ${\mathbf{a}}$ and ${\mathbf{b}}$ over $\mathbb{U}$, their aperiodic cross-correlation function (ACCF) is defined as
	\begin{equation}\label{defi_ACCF}
	\rho_{\mathbf{a},\mathbf{b}}(\tau):= \left \{
	\begin{array}{cl}
	\sum\limits_{k=0}^{N-1-\tau}{a_kb_{k+\tau}},&~~0\leq \tau \leq N-1;\\
	\sum\limits_{k=0}^{N-1-\tau}{a_{k+\tau}b_k},&~~-(N-1)\leq \tau \leq -1;\\
	0,& ~~\mid \tau \mid \geq N.
	\end{array}
	\right .
	\end{equation}
	When $\mathbf{a} = \mathbf{b}$, $\rho_{\mathbf{a},\mathbf{b}}(\tau)$ is called aperiodic auto-correlation function (AACF) of $\mathbf{a}$ and  will be denoted as $\rho_{\mathbf{a}}(\tau)$.
	
	\subsection{Introduction to OB-ZCPs}
	Before we proceed further, it is worthy to point out that, sometimes, we write two row sequences (of identical sequence length) $\mathbf{a}$ and $\mathbf{b}$ as $(\mathbf{a};\mathbf{b})$, a two-dimensional matrix obtained from vertical concatenation of $\mathbf{a}$ and $\mathbf{b}$.
	
	\vspace{0.1in}
	\begin{definition}[Type-I binary Z-complementary pair \cite{zilong_obzcp}]
		Sequences $\mathbf{a}$ and $\mathbf{b}$ of length-$N$ are said to be a Type-I ZCP with ZCZ width $Z$ if and only if
		\begin{equation}
		\rho_{\mathbf{a}}(\tau)+\rho_{\mathbf{b}}(\tau)=0,~~\text{for all}~1\leq \tau \leq Z-1.
		\end{equation}
		If $Z=N$, $(\mathbf{a};\mathbf{b})$ reduces to a GCP \cite{golay1}.
	\end{definition}
	\vspace{0.1in}
	\begin{definition}[Type-II binary Z-complementary pair \cite{zilong_obzcp}]
		Sequences $\mathbf{a}$ and $\mathbf{b}$ of length-$N$ are said to be a Type-II ZCP with ZCZ width $Z$ if and only if
		\begin{equation}
		\rho_{\mathbf{a}}(\tau)+\rho_{\mathbf{b}}(\tau)=0,~~\text{for all} ~(N-Z+1)\leq \tau \leq  N-1.
		\end{equation}
	\end{definition}
	\vspace{0.1in}
	\begin{definition}[OB-ZCP]
A binary ZCP, Type-I or Type-II, is called an OB-ZCP if its sequence length $N$ is odd.
	\end{definition}
	\vspace{0.1in}
	\begin{lemma}\cite{zilong_obzcp}
		Every Type-I (or Type-II) OB-ZCP of length $N$ (odd) has a maximum ZCZ width $(N+1)/2$, i.e.,
		\begin{equation}
		Z \leq (N+1)/2.
		\end{equation}
	\end{lemma}
	\vspace{0.1in}
	\begin{definition}\cite{zilong_obzcp}
		An OB-ZCP (Type-I or Type-II) is said to be \textit{Z-optimal} if $Z=(N+1)/2$.
	\end{definition}
	\vspace{0.1in}
	\begin{lemma}\cite{zilong_obzcp}
		The magnitude of each out-of-zone aperiodic auto-correlation sum for a \textit{Z-optimal} Type-I OB-ZCP $(\mathbf{a};\mathbf{b})$ of length $N$ is lower bounded by $2$, i.e.,
		\begin{equation}
		\mid \rho_{\mathbf{a}}(\tau)+\rho_{\mathbf{b}}(\tau) \mid \geq 2, ~ \text{ for any } (N+1)/2 \leq \tau \leq (N-1).
		\end{equation}
	\end{lemma}
	\begin{lemma}\cite{zilong_obzcp}
		The magnitude of each out-of-zone aperiodic auto-correlation sum for a \textit{Z-optimal} Type-II OB-ZCP $(\mathbf{c};\mathbf{d})$ of length $N$ is lower bounded by $2$, i.e.,
		\begin{equation}
		\mid \rho_{\mathbf{c}}(\tau)+\rho_{\mathbf{d}}(\tau) \mid \geq 2, ~ \text{ for any } 1 \leq \tau \leq (N-1)/2.
		\end{equation}
	\end{lemma}
	\vspace{0.1in}
	\begin{definition}[\textit{Optimal} OB-ZCP \cite{zilong_obzcp}]
		An OB-ZCP (Type-I or Type-II) is said to be \textit{optimal} if it is \textit{Z-optimal} and every out-of-zone AACS takes the minimum magnitude value of 2.
	\end{definition}
	\vspace{0.1in}
	\begin{example}
		Let
		\begin{equation}
		\begin{array}{lcl}
		\mathbf{a}& =& (+++-++-++),\\
		\mathbf{b}& =& (+++---+-+).
		\end{array}
		\end{equation}
		$(\mathbf{a};\mathbf{b})$ is a length-$9$ Type-I optimal OB-ZCP with a ZCZ width of $5$ because
		\begin{equation}
		\begin{array}{ccl}
		\Bigl ( \mid \rho_{\mathbf{a}}(\tau)+\rho_{\mathbf{b}}(\tau) \mid \Bigl )_{\tau=0}^{8} & = & (18,0,0,0,0,2,2,2,2).
		\end{array}
		\end{equation}
	\end{example}
	\vspace{0.1in}
	\begin{example}
		Let
		\begin{equation}
		\begin{array}{lcl}
		\mathbf{c}& =& (-+++-+-++),\\
		\mathbf{d}& =& (-+++--+--).
		\end{array}
		\end{equation}
		$(\mathbf{c};\mathbf{d})$ is a length-$9$ Type-II optimal OB-ZCP with a ZCZ width of $5$ because
		\begin{equation}
		\begin{array}{ccl}
		\Bigl ( \mid \rho_{\mathbf{c}}(\tau)+\rho_{\mathbf{d}}(\tau) \mid \Bigl )_{\tau=0}^{8} & = & (18,2,2,2,2,0,0,0,0).
		\end{array}
		\end{equation}
	\end{example}
	
	\subsection{Properties of Binary GCPs and Insertion Operator}
	In this subsection, we present some properties relevant to binary GCPs and insertion operator. These properties will be useful in the proof of new \textit{optimal} OB-ZCPs in Section III.
	\vspace{0.05in}
	\begin{lemma}\cite{Borwein03}\label{prop_quad}
		Let $\mathcal{G}\equiv (\mathbf{a};\mathbf{b})$ be a binary GCP of length $N$ given by
		\begin{equation}
		\mathcal{G}=\left ( \begin{matrix}
		a_0,a_1,\dots,a_{N-1}\\
		b_0,b_1,\dots,b_{N-1}
		\end{matrix} \right).
		\end{equation}
		Then,
		\begin{equation}
		a_i+a_{N-1-i}+b_i+b_{N-1-i}=\pm 2, ~ ~ 0 \leq i < N/2.
		\end{equation}
		This shows that if the $i$-th column vector of $\mathcal{G}$ is comprised of two identical binary elements, then the $(N-i)$-{th} column vector should consist of two elements with opposite signs, and vice versa.
	\end{lemma}

	\vspace{0.1in}
	
	Binary GCPs can be obtained from sequence operations based on several kernels\footnote{Sequence pairs which cannot be obtained from any shorter pairs of sequences.} of lengths $2,~ 10$ and $26$. These kernels \cite{Borwein03} are listed in Table \ref{table_new} and will be useful in Section III.

	\renewcommand{\arraystretch}{0.6}
	\begin{table}
		\small
		\centering
		\tabcolsep=0.11cm
		\caption{\cite{Borwein03} GCP Kernels of Lengths $2$, $10$ and $26$. \label{table_new}}
		\resizebox{\columnwidth}{!}{
			
			\begin{tabular}{|c||c||c|}
				\hline
				$N$ & $\left( \begin{matrix}
				\mathbf{a}\\
				\mathbf{b}
				\end{matrix} \right)$ & Notation   \\ \hline \hline
				$2$  & $\left( \begin{matrix}
				++\\
				+-
				\end{matrix} \right) $ & $K_2$  \\  \hline
				$10$ & $ \left ( \begin{matrix}
				++-+-+--++\\
				++-+++++--
				\end{matrix} \right) $ & $K_{10}$  \\  \hline
				$26$ & $\left ( \begin{matrix}
				++++-++--+-+-+--+-+++--+++\\
				++++-++--+-+++++-+---++---
				\end{matrix} \right) $ & $K_{26}$  \\  \hline
			\end{tabular}
		}
	\end{table}
	\vspace{0.1in}
	\begin{definition}[Insertion Operator \cite{zilong_obzcp}] Let $\mathbf{a}=(a_0,a_1,\cdots,a_{N-1})$ be a sequence of length $N$. Define $\mathcal{I}(\mathbf{a},r,x)$, where $r\in\{0,1,\cdots,N\}$, an insertion operator which generates a length-$(N+1)$ sequence with element $x$, as follows.
		\begin{equation}\label{definsertion}
		\mathcal{I}(\mathbf{a},r,x) =
		\begin{cases}
		(x,a_0,a_1,\dots,a_{N-1}), & \text{if }r=0; \\
		(a_0,a_1,\dots,a_{N-1},x), & \text{if }r=N; \\
		(a_0,a_1,\dots,a_{r-1},x,a_r,\cdots,a_{N-1}), & \hspace{-0.1in}0<r<N. \\
		\end{cases}
		\end{equation}
	\end{definition}
	
	\vspace{0.1in}
	\begin{lemma}\label{aacfinsertion}
		For a binary sequence $\mathbf{a}$ of length $N$ (even), denote by $\mathbf{a}^1$ and $\mathbf{a}^2$ the first and second halves of $\mathbf{a}$, respectively, i.e.,
		\begin{equation}\label{seq_rep}
		\begin{array}{lcl}
		\mathbf{a}^1& =& (a_0,~ a_1,~ \dots,  ~ a_{N/2-1}),\\
		\mathbf{a}^2& =& (a_{N/2},~ a_{N/2+1},~ \dots,  ~ a_{N-1}).
		\end{array}
		\end{equation}
		The AACF of $\mathcal{I}(\mathbf{a},r,x)$ (where $x \in \mathbb{U}$) is shown in (\ref{eqnfront}), (\ref{eqnmiddle}) and (\ref{eqnend}) for different values of $r$.
	\end{lemma}
	\noindent 1), If $r=0$:
	\begin{equation} \label{eqnfront}
	\rho_{\mathcal{I}(\mathbf{a},r,x)}(\tau) =
	xa_{\tau-1}+\rho_{\mathbf{a}}(\tau), ~~~~~~ \text{if } 0<\tau \leq N.
	\end{equation}
	\noindent 2), If $r=\frac{N}{2}$:
	\begin{equation} \label{eqnmiddle}
	\begin{split}
	& \rho_{\mathcal{I}(\mathbf{a},r,x)}(\tau) = \\
	& \begin{cases}
	\rho_{\mathbf{a}^1}(\tau)+xa_{r-\tau}+\rho_{\mathbf{a}^2,\mathbf{a}^1}(r-\tau+1)+ \\  \hfill \quad xa_{r+\tau-1}+\rho_{\mathbf{a}^2}(\tau), & \text{if } \tau < r \\
	xa_{r-\tau}+\rho_{\mathbf{a}^2,\mathbf{a}^1}(r-\tau+1)+xa_{r+\tau-1}, & \text{if } \tau= r \\
	\rho_{\mathbf{a}^2,\mathbf{a}^1}(r-\tau+1), & \text{if } \tau > r
	\end{cases}
	\end{split}
	\end{equation}
	\noindent 3), If $r=N$:
	\begin{equation} \label{eqnend}
	\rho_{\mathcal{I}(\mathbf{a},r,x)}(\tau) =
	xa_{r-\tau}+\rho_{\mathbf{a}}(\tau), ~~~~~~ \text{if } 0<\tau \leq N.
	\end{equation}

	\section{CONSTRUCTIONS OF \textit{optimal} OB-ZCPs}
	In this section, we present systematic constructions of new sets of \textit{optimal} OB-ZCPs by employing insertion method to binary GCPs having \textit{generic} lengths of $2^\alpha 10^\beta 26^\gamma$. In the next subsection, we first recall Turyn's construction method for binary GCPs and then reveal several intrinsic structure properties pertinent to these GCPs. These properties play a fundamental role in proving these newly constructed \textit{optimal} OB-ZCPs.
	\subsection{Intrinsic Structure  Properties of binary GCPs from Turyn's Construction}
	 Our objective is to uncover several intrinsic structure properties pertinent to the columns of binary GCPs which are generated from Turyn's construction.
	\vspace{0.1in}
	\begin{lemma}[Turyn's Method \cite{Turyn74}]\label{thm_turyn}
		Let $\mathcal{A}\equiv(\mathbf{a};\mathbf{b})$ and $\mathcal{B}\equiv (\mathbf{c};\mathbf{d})$ be binary GCPs of lengths $N$ and $M$ respectively and denote $\mathcal{A}$ as the $1st$ pair and $\mathcal{B}$ as the $2nd$ pair. Then $(\mathbf{e};\mathbf{f})\triangleq Turyn(\mathcal{A},\mathcal{B})$ is a GCP of length-$MN$ where,
		\begin{equation} \label{turyn_1st_form}
		\begin{split}
		\mathbf{e} & =\mathbf{c} \otimes \left(\mathbf{a}+\mathbf{b} \right) / 2 - \overleftarrow{\mathbf{d}} \otimes \left(\mathbf{b}-\mathbf{a} \right) / 2,  \\ \mathbf{f} & =\mathbf{d} \otimes \left( \mathbf{a}+\mathbf{b} \right) / 2 + \overleftarrow{\mathbf{c}} \otimes  \left( \mathbf{b}-\mathbf{a} \right) / 2.
		\end{split}
		\end{equation}
	\end{lemma}
	\vspace{0.1in}

\vspace{0.1in}

The $i$-th elements of $\mathbf{e}$ and $\mathbf{f}$ are given in (\ref{new1}).
\begin{figure*}
	\begin{equation}\label{new1}
	\begin{split}
	e_i=\frac{a_{i\text{ mod }N}}{2}\left(c_{\left \lfloor \frac{i}{N} \right \rfloor}+d_{M-\left \lfloor \frac{i}{N} \right \rfloor-1}\right)+\frac{b_{i\text{ mod }N}}{2}\left(c_{\left \lfloor \frac{i}{N} \right \rfloor}-d_{M-\left \lfloor \frac{i}{N} \right \rfloor-1}\right),\\
	f_i=\frac{a_{i\text{ mod }N}}{2}\left(d_{\left \lfloor \frac{i}{N} \right \rfloor}-c_{M-\left \lfloor \frac{i}{N} \right \rfloor-1}\right)+\frac{b_{i\text{ mod }N}}{2}\left(d_{\left \lfloor \frac{i}{N} \right \rfloor}+c_{M-\left \lfloor \frac{i}{N} \right \rfloor-1}\right).
	\end{split}
	\end{equation}
\end{figure*}
\vspace{0.1in}

In our construction, we have fixed $\mathcal{A}\equiv (\mathbf{a};\mathbf{b})$ to be a kernel GCP, given in Table I. Then we have the following three cases.
\begin{enumerate}
	\item When $\mathcal{A}=K_2$, we have $a_0 = b_0$ and $a_1 = -b_1$.
	\item When $\mathcal{A}=K_{10}$, we have $a_i = b_i$ for $i\in\{0,1,2,3,5\}$, $a_i = -b_i$ for $i\in\{4,6,7,8,9\}$.
	\item When $\mathcal{A}=K_{26}$, we have $a_i = b_i$ for $i\in\{0,1,\cdots,11,13\}$, $a_i = -b_i$ for $i\in\{12,14,15,\cdots,25\}$.
\end{enumerate}

\vspace{0.1in}

From (\ref{new1}), we have
\begin{equation}\label{new2}
\begin{split}
e_i&=\begin{cases}
a_{i\text{ mod }N}c_{\left \lfloor \frac{i}{N} \right \rfloor} & \text{ if }a_{i\text{ mod }N}=b_{i\text{ mod }N},\\
a_{i\text{ mod }N}d_{M-\left \lfloor \frac{i}{N} \right \rfloor-1} & \text{ if }a_{i\text{ mod }N}=-b_{i\text{ mod }N},
\end{cases}\\
f_i&=\begin{cases}
a_{i\text{ mod }N}d_{\left \lfloor \frac{i}{N} \right \rfloor} & \text{ if }a_{i\text{ mod }N}=b_{i\text{ mod }N},\\
-a_{i\text{ mod }N}c_{M-\left \lfloor \frac{i}{N} \right \rfloor-1} & \text{ if }a_{i\text{ mod }N}=-b_{i\text{ mod }N}.
\end{cases}
\end{split}
\end{equation}

Since $(\mathbf{c};\mathbf{d})$ is a binary GCP, from \textit{Lemma} \ref{prop_quad}, we assert that if $c_{\left \lfloor \frac{i}{N} \right \rfloor}$ and $d_{\left \lfloor \frac{i}{N} \right \rfloor}$ are of identical signs, $c_{M-\left \lfloor \frac{i}{N} \right \rfloor-1}$ and $d_{M-\left \lfloor \frac{i}{N} \right \rfloor-1}$ must be of different signs. Therefore, from (\ref{new2}), we see that the elements $e_i$ and $f_i$ are of identical signs, provided that $c_{\left \lfloor \frac{i}{N} \right \rfloor}$ and $d_{\left \lfloor \frac{i}{N} \right \rfloor}$ are of identical signs. Specifically, provided that $c_{\left \lfloor \frac{i}{N} \right \rfloor}=d_{\left \lfloor \frac{i}{N} \right \rfloor}$, the columns of length-$N$ sub-sequence pair
\begin{equation}\label{new3}
\left ( \begin{matrix}
e_{kN+0},e_{kN+1},\dots,e_{kN+N-1}\\
f_{kN+0},f_{kN+1},\dots,f_{kN+N-1}
\end{matrix} \right)
\end{equation}
have identical signs in each column, where $0\leq k<M$.

On the other hand, if $c_{\left \lfloor \frac{i}{N} \right \rfloor}\neq d_{\left \lfloor \frac{i}{N} \right \rfloor}$, \textit{Lemma} \ref{prop_quad} shows that $c_{M-\left \lfloor \frac{i}{N} \right \rfloor-1}$ and $d_{M-\left \lfloor \frac{i}{N} \right \rfloor-1}$ are of identical signs, and therefore from (\ref{new2}) the elements $e_i$ and $f_i$ are of different signs. That is, every column of (\ref{new3}) will have different signs.
	
\vspace{0.1in}
	
	Based on the above analysis, we present the following theorem.
	
	\vspace{0.1in}
	\begin{theorem}\label{turyn_func}
		Let $\mathcal{A} \equiv (\mathbf{a};\mathbf{b})$ be a binary GCP kernel $K_N $ where $N\in \{2,10,26\}, \mathcal{B} \equiv (\mathbf{c};\mathbf{d})$ be a GCP of length $M$ and $(\mathbf{e};\mathbf{f})=Turyn(\mathcal{A},\mathcal{B})$. If the $i$-th column of $\mathcal{B}$ have elements with same sign, then we have $e_t=f_t, ~Ni \leq t < N(i+1)$. In other words, each column of $(\mathbf{e};\mathbf{f})$ with column  indices ranging from $Ni$ to $Ni+(N-1)$ will have elements with identical signs. If the $i$-th column of $\mathcal{B}$ have elements with different signs, then we have $e_t=-f_t, ~Ni \leq t < N(i+1)$, i.e., each column of $(\mathbf{e};\mathbf{f})$ with column indices ranging from $Ni$ to $Ni+(N-1)$ will have elements with different signs.
	\end{theorem}
	\vspace{0.1in}
	\begin{example}\label{examle3_tag}
		Let $\mathcal{A}=K_2$ and $\mathcal{B}\equiv(\mathbf{c};\mathbf{d})$ be a GCP of length $4$ given as follows.
		\begin{equation}
		\left( \begin{matrix}
		\mathbf{c} \\
		\mathbf{d}
		\end{matrix} \right) = \left( \begin{matrix}
		\textcolor{blue}{+}\textcolor{red}{+}\textcolor{blue}{+}\textcolor{red}{-}\\
		\textcolor{blue}{+}\textcolor{red}{-}\textcolor{blue}{+}\textcolor{red}{+}\end{matrix} \right)
		\end{equation}
%
%

Then $(\mathbf{e};\mathbf{f})= Turyn(\mathcal{A},\mathcal{B})$ is given below.

		\begin{equation}
		\left( \begin{matrix}
		\mathbf{e} \\
		\mathbf{f}
		\end{matrix} \right) = \left( \begin{matrix}
		\textcolor{blue}{++}\textcolor{red}{++}\textcolor{blue}{+-}\textcolor{red}{-+}\\
		\textcolor{blue}{++}\textcolor{red}{--}\textcolor{blue}{+-}\textcolor{red}{+-}\end{matrix} \right)
		\end{equation}
		Note that 1), each of first two columns of $(\mathbf{e};\mathbf{f})$, i.e.,  $\left (\begin{smallmatrix}\textcolor{blue}{++}\\\textcolor{blue}{++}\end{smallmatrix}\right)$, has elements of identical signs because the first column of $\mathcal{B}$, i.e., $\left( \begin{smallmatrix}\textcolor{blue}{+}\\\textcolor{blue}{+}\end{smallmatrix}\right)$ has identical sign, 2), each of the next two columns of $(\mathbf{e};\mathbf{f})$, i.e., $\left(\begin{smallmatrix}\textcolor{red}{++}\\\textcolor{red}{--}\end{smallmatrix}\right)$ has elements of opposite signs because the second column of $\mathcal{B}$, i.e., $\left( \begin{smallmatrix}\textcolor{red}{+}\\\textcolor{red}{-}\end{smallmatrix}\right)$ has different signs. Continuing this check to the last column of $(\mathbf{e};\mathbf{f})$, one can verify \textit{Theorem} \ref{turyn_func}.
	\end{example}
	\vspace{0.1in}
	Based on \textit{Theorem} \ref{turyn_func}, we can easily obtain the following corollaries. These corollaries lead to certain GCPs which will be employed for the constructions of our proposed OB-ZCPs in the next section. Specifically,
  \begin{itemize}
	\item Corollary 2 will be used in Construction 1, Construction 3, and Construction 5.
	\item Corollary 3 will be used in Construction 2 and Construction 4.
	\item Corollary 4 will be used in Cases 3 and 6 of Table III and Table IV.
\end{itemize}

	\vspace{0.1in}
	\begin{corollary}\label{tur_cor}
		Consider $\mathcal{A},\mathcal{B},(\mathbf{e};\mathbf{f})$ described in \textit{Theorem} \ref{turyn_func}. If there are consecutive $t$ columns of $\mathcal{B}$ having elements with identical signs in each column, starting from the $i$-th column, then $(\mathbf{e};\mathbf{f})$ will have consecutive $tN$ columns consisting of elements with identical signs in each column, starting from the $Ni$-th column.
	\end{corollary}
	\vspace{0.1in}
	
	\begin{corollary}\label{fac_2_tur}
			Let $(\mathbf{e};\mathbf{f})$ be a GCP of length $2^\alpha M$, constructed iteratively by employing Turyn's method on $K_2,~ K_{10} \text{ and }K_{26}$ as follows:
			\begin{equation}\label{fac_2_tur_equ}
			\begin{split}
			& (\mathbf{e}_0;\mathbf{f}_0)=K_2, \\
			& (\mathbf{e}_i;\mathbf{f}_i)=Turyn(\mathcal{A},(\mathbf{e}_{i-1};\mathbf{f}_{i-1})), ~ \mathcal{A}=K_2,~K_{10} \text{ or } K_{26},
			\end{split}
			\end{equation}
			where $M=10^\beta 26^\gamma$ and $\alpha,~\beta, \text{ and } \gamma$ are non-negative integers and $\alpha \geq 1$. Then the first $2^{\alpha -1} M$ columns of $(\mathbf{e};\mathbf{f})$ will have elements with identical sign in each column.
	\end{corollary}
	\vspace{0.1in}
	\begin{example}\label{ex_2_10}
		Let $(\mathbf{e};\mathbf{f})$ be a GCP of length $20$, constructed via Turyn's method by setting $\mathcal{A}=K_{10}$ for $i=1$ in \textit{Corollary} \ref{fac_2_tur}. The first $(2^0 \times 10^1)=10$ columns of $\mathcal{C}$ has entries with identical signs in each column.
		\begin{equation}\label{examp2_10}
		\left( \begin{matrix}
		\mathbf{e} \\
		\mathbf{f}
		\end{matrix} \right)= \left ( \begin{matrix}
		\textcolor{blue}{++-+-+--++}--+-----++\\
		\textcolor{blue}{++-+-+--++}++-+++++--
		\end{matrix} \right)
		\end{equation}
	\end{example}
	\vspace{0.1in}
	\begin{corollary}\label{tur_th_pow}
			Let $(\mathbf{e};\mathbf{f})$ be a GCP of length $N^p$, constructed recursively by employing Turyn's method on $K_N$, where $N\in \{ 2,10,26\}$ and $p$ is a non-negative integer. Also suppose there are $t$ consecutive columns of $K_N$, $N\in \{ 2,10,26\}$, having elements with identical signs (or different signs) in each column, starting from the $i$-th column index. Then, the $tN^{p-1}$ consecutive columns of $(\mathbf{e};\mathbf{f})$ will have elements with identical sign (or different sign) in each column, starting from the $iN^{p-1}$-th column.
	\end{corollary}
	\vspace{0.1in}
	\begin{example} \label{example_3}
		Let $(\mathbf{e};\mathbf{f})=Turyn(K_{10},K_{10})$ be a GCP of length $100$. The first $(4 \times 10^{2-1})=40$ columns of $(\mathbf{e};\mathbf{f})$ has entries with same sign in each column.
		\begin{equation} \label{gcp100_1}
		\left( \begin{matrix}
		\mathbf{e} \\
		\mathbf{f}
		\end{matrix} \right)= \left ( A \Vert B \Vert C \Vert D \right),
		\end{equation}
		where $A,B,C$ and $D$ are the following sequence pairs of length $25$,
		\begin{equation}\label{gcp100}
		\begin{split}
		& A= \left ( \begin{smallmatrix}
		\textcolor{blue}{++-+++++--++-+++++----+--} \\
		\textcolor{blue}{++-+++++--++-+++++----+--}
		\end{smallmatrix} \right), \\
		& B= \left ( \begin{smallmatrix}
		\textcolor{blue}{---++++-+-+--++}--+-----++\\
		\textcolor{blue}{---++++-+-+--++}++-+++++--
		\end{smallmatrix} \right),\\
		& C= \left ( \begin{smallmatrix}
		++-+-+--++--+-----++--+-+\\
		++-+-+--++++-+++++--++-+-
		\end{smallmatrix} \right), \\
		& D= \left ( \begin{smallmatrix}
		-++--++-+-+--++++-+-+--++\\
		+--++--+-+-++----+-+-++--
		\end{smallmatrix} \right). \\
		\end{split}
		\end{equation}
		Similar to \textit{Example \ref{examle3_tag}}, \textit{Corollary \ref{tur_th_pow}} can be verified.
	\end{example}
	
	\vspace{0.1in}
	
	\begin{corollary}\label{th_pow_1026_1}
		Let $(\mathbf{e};\mathbf{f})$ be a GCP of length $10^\beta 26^\gamma$, constructed iteratively by employing Turyn's method on $K_{10} \text{ and }K_{26}$ as follows:
		\begin{equation}\label{fac_2_tur_equ2}
		\begin{split}
		& (\mathbf{e}_0;\mathbf{f}_0)=K_{26}, \\
		& (\mathbf{e}_i;\mathbf{f}_i)=Turyn(\mathcal{A},(\mathbf{e}_{i-1};\mathbf{f}_{i-1})), ~ \mathcal{A}=K_{10} \text{ or } K_{26},
		\end{split}
		\end{equation}
		where $\beta \text{ and } \gamma$ are non-negative integers. Then the first $12\times 26^{\gamma -1} 10^\beta$ columns of $(\mathbf{e};\mathbf{f})$ will have elements with identical signs in each column.
	\end{corollary}
	\vspace{0.1in}
	\begin{remark}
		Setting $(\mathbf{e}_0;\mathbf{f}_0)$ to $K_{10}$, we assert that the first $4\times 10^{\beta-1} 26^\gamma$ columns have identical signs in each column. The OB-ZCPs constructed using these GCPs, however, will have less ZCZ widths, compared to the OB-ZCPs by GCPs given in \textit{Corollary} \ref{th_pow_1026_1}.
	\end{remark}
	\vspace{0.1in}
	\subsection{Proposed Constructions of New Optimal Type-I and Type-II OB-ZCPs}
	For ease of presentation, from hereon, we assume the GCPs, except the kernel GCPs, used in this paper are obtained from any construction in \textit{Corollaries} \ref{fac_2_tur}-\ref{th_pow_1026_1}. The following lemma gives the AACSs at different time-shifts when $x,y \in \mathbb{U}$ are inserted at the front position of the sequences $\mathbf{a}$ and $\mathbf{b}$, respectively.
	\vspace{0.1in}
	\begin{lemma}\label{th_front}
		Let ($\mathbf{a};\mathbf{b}$) be GCP of length $N$. If $\mathbf{e}={\mathcal{I}(\mathbf{a},0,x)}$ and $\mathbf{f}={\mathcal{I}(\mathbf{b},0,y)}$, where $x,y \in \mathbb{U}$, then it satisfies the following:
		\begin{equation}
		\rho_{\mathbf{e}}(\tau)+\rho_{\mathbf{f}}(\tau)=xa_{\tau-1}+yb_{\tau-1}, ~~~ 1\leq \tau <N.
		\end{equation}
	\end{lemma}
	\vspace{0.1in}
	\begin{IEEEproof}
		From (\ref{eqnfront}), for $1\leq \tau <N $. we get,
		\begin{equation}\label{magnitude}
		\begin{split}
		\rho_{\mathbf{e}}(\tau)+\rho_{\mathbf{f}}(\tau) &= xa_{\tau-1}+yb_{\tau-1}+\rho_{\mathbf{a}}(\tau)+\rho_{\mathbf{b}}(\tau) \\
		& = xa_{\tau-1}+yb_{\tau-1}
		\end{split}
		\end{equation}
	\end{IEEEproof}
	\vspace{0.1in}
	Note that (\ref{magnitude}) can be further reduced to
	\begin{equation}\label{magnitude2}
	\rho_{\mathbf{e}}(\tau)+\rho_{\mathbf{f}}(\tau)=
	\begin{cases}
	0,~~ & \text{if}~xa_{\tau-1} \cdot yb_{\tau-1} = -1, \\
	\pm 2,~~ & \text{if}~xa_{\tau-1} \cdot yb_{\tau-1} = 1.
	\end{cases}.
	\end{equation}
	We now present the following construction by recalling {\textit{Corollary} \ref{fac_2_tur}.}
	\vspace{0.1in}
	
	\begin{construction}\label{th_pow_2_mag}
		Let $(\mathbf{a};\mathbf{b})$ be a GCP generated by (\ref{fac_2_tur_equ}) which has length $N=2^\alpha M$, where $M=10^\beta 26^\gamma$, $\alpha,\beta, \gamma$ are non-negative integers and $\alpha \geq 1$. If $x,y \in \mathbb{U}$ and $\mathbf{e}={\mathcal{I}(\mathbf{a},0,x)}$, $\mathbf{f}={\mathcal{I}(\mathbf{b},0,y)}$, Then we have the following constructions:
		\begin{itemize}
			\item $(\mathbf{e};\mathbf{f})$ is a Type-I \textit{optimal} OB-ZCP with ZCZ width of $2^{\alpha -1} M +1$ satisfying the following equation.
			\begin{equation}\label{type1_2}
			\mid\rho_{\mathbf{e}}(\tau)+\rho_{\mathbf{f}}(\tau)\mid=
			\begin{cases}
			0,& \text{if }  0 < \tau \leq 2^{\alpha -1} M, \\
			2,& \text{if } 2^{\alpha -1} M < \tau <2^\alpha M.
			\end{cases}
			\end{equation}
			when $x$ and $y$ are of different signs.
			\item $(\mathbf{e};\mathbf{f})$ is a Type-II \textit{optimal} OB-ZCP with ZCZ width of $2^{\alpha -1} M +1$, i.e.,
			\begin{equation}\label{type2_2}
			\mid\rho_{\mathbf{e}}(\tau)+\rho_{\mathbf{f}}(\tau)\mid=
			\begin{cases}
			2,& \text{if }  0 < \tau \leq 2^{\alpha -1} M, \\
			0,& \text{if } 2^{\alpha -1} M < \tau <2^\alpha M.
			\end{cases}
			\end{equation}
			when $x$ and $y$ are of identical signs.
		\end{itemize}
	\end{construction}

	\vspace{0.1in}
	\begin{IEEEproof}
		By \textit{Corollary} \ref{fac_2_tur}, we have $a_i=b_i \text{ for } 0< i \leq 2^{\alpha -1} M$. Also, by \textit{Lemma} \ref{prop_quad}, we have $a_i=-b_i,\text{ where } 2^{\alpha -1}M< i \leq 2^\alpha M$. Based on (\ref{magnitude2}), we complete the first proof by considering $x$ and $y$ with different signs ($x,y \in \mathbb{U}$) and which are inserted at the front position of $\mathbf{a}$ and $\mathbf{b}$, respectively. The second proof follows similarly, considering $x$ and $y$ with identical signs.
	\end{IEEEproof}
	\vspace{0.1in}
	\begin{remark}
		In \textit{Construction} \ref{th_pow_2_mag}, if $\beta=0,~\gamma=0$ then it leads to certain optimal OB-ZCPs given in \textit{Theorem} $4$ of \cite{zilong_obzcp}.
	\end{remark}
	\vspace{0.1in}
	\begin{example}\label{ex_2_10_z1}
		Let $(\mathbf{a};\mathbf{b})$ be the GCP of length $20$ shown in (\ref{examp2_10}). If $x=1$, $y=-1$, then $\mathbf{e}\equiv \mathcal{I}(\mathbf{a},0,x)$, $\mathbf{f}\equiv \mathcal{I}(\mathbf{b},0,y)$ is
		\begin{equation}
		\left( \begin{matrix}
		\mathbf{e} \\
		\mathbf{f}
		\end{matrix} \right) = \left( \begin{matrix}
		\textcolor{green}{+}\textcolor{blue}{++-+-+--++}--+-----++\\
		\textcolor{green}{-}\textcolor{blue}{++-+-+--++}++-+++++--
		\end{matrix} \right)
		\end{equation}
		One can readily show that $(\mathbf{e};\mathbf{f})$ is a length-$21$ \textit{optimal} Type-I OB-ZCP with a ZCZ width of $11$ because
		\begin{equation}
		\begin{array}{ccl}
		\Bigl ( \mid \rho_{\mathbf{e}}(\tau)+\rho_{\mathbf{f}}(\tau) \mid \Bigl )_{\tau=0}^{20} & = & (42,\mathbf{0}_{10},\mathbf{2}_{10}).
		\end{array}
		\end{equation}
	\end{example}
	\vspace{0.1in}

	\begin{example}\label{ex_2_10_z}
		Let $(\mathbf{a};\mathbf{b})$ be the GCP of length 20 shown in (\ref{examp2_10}).
		If $x=1$, $y=1$, then $\mathbf{e}\equiv \mathcal{I}(\mathbf{a},0,x)$, $\mathbf{f}\equiv \mathcal{I}(\mathbf{b},0,y)$ is
	{\color{black}	\begin{equation}
		\left( \begin{matrix}
		\mathbf{e} \\
		\mathbf{f}
		\end{matrix} \right) = \left( \begin{matrix}
		\textcolor{green}{+}\textcolor{blue}{++-+-+--++}--+-----++\\
		\textcolor{green}{+}\textcolor{blue}{++-+-+--++}++-+++++--
		\end{matrix} \right)
		\end{equation} }
		Here, $(\mathbf{e};\mathbf{f})$ is a length-$21$ \textit{optimal} Type-II OB-ZCP with a ZCZ width of $11$ because
		\begin{equation}
		\begin{array}{ccl}
		\Bigl ( \mid \rho_{\mathbf{e}}(\tau)+\rho_{\mathbf{f}}(\tau) \mid \Bigl )_{\tau=0}^{20} & = & (42,\mathbf{2}_{10},\mathbf{0}_{10}).
		\end{array}
		\end{equation}
	\end{example}
\vspace{0.1in}
	Recalling \textit{Corollary} \ref{tur_th_pow}, we obtain the following construction.
	\vspace{0.1in}
	
	\begin{construction}\label{th_pow_10_26}
		Let $(\mathbf{a};\mathbf{b})$ be a GCP generated by \textit{Corollary} \ref{tur_th_pow} and of length $N^p$, where $N=10 \text{ or } 26$ and $p$ is a non-negative integer. If $x,y \in \mathbb{U}$ and $\mathbf{e}={\mathcal{I}(\mathbf{a},0,x)}$, $\mathbf{f}={\mathcal{I}(\mathbf{b},0,y)}$, then we have the following constructions:
		\begin{itemize}
		 \item $(\mathbf{e};\mathbf{f})$ is a Type-I OB-ZCP with ZCZ width of $(i_0+t_0)N^{p-1} +1$ satisfying the following equation.
		\begin{equation}
		\begin{split}
		& \mid\rho_{\mathbf{e}}(\tau)+\rho_{\mathbf{f}}(\tau)\mid \\ &=
		\begin{cases}
		0,& \text{if }  i_0N^{p-1} < \tau \leq (i_0+t_0)N^{p-1}, \\
		2,& \text{if } i_1N^{p-1} < \tau \leq(i_1+t_1)N^{p-1} ,\\
		0,& \text{if } i_2N^{p-1} < \tau \leq (i_2+t_2)N^{p-1} , \\
		2,& \text{if } i_3N^{p-1} < \tau <(i_3+t_3)N^{p-1}=N^p,
		\end{cases}
		\end{split}
		\end{equation}
		when $x$ and $y$ are of different signs.
		\item $(\mathbf{e};\mathbf{f})$ is a Type-II OB-ZCP with ZCZ width of $(i_0+t_0)N^{p-1} +1$, i.e.,
		\begin{equation}
		\begin{split}
		& \mid\rho_{\mathbf{e}}(\tau)+\rho_{\mathbf{f}}(\tau)\mid \\ & =
		\begin{cases}
		2,& \text{if }  i_0N^{p-1} < \tau \leq (i_0+t_0)N^{p-1}, \\
		0,& \text{if } i_1N^{p-1} < \tau \leq(i_1+t_1)N^{p-1} ,\\
		2,& \text{if } i_2N^{p-1} < \tau \leq (i_2+t_2)N^{p-1} , \\
		0,& \text{if } i_3N^{p-1} < \tau <(i_3+t_3)N^{p-1} =N^p,
		\end{cases}
		\end{split}
		\end{equation}
		when $x$ and $y$ are identical signs.
		\end{itemize}
		Here $t_n$ is the number of consecutive columns of $K_N$, each having elements with identical/different sign (depending on the value of $t_n$), starting from the $i_n$-th column. The values of $(i_n,t_n)$ for $K_{10}$ and $K_{26}$ are given in Table \ref{table_new45}.
	\end{construction}

	\vspace{0.1in}
	\begin{IEEEproof}
		Since $(\mathbf{a};\mathbf{b})$ is generated using $K_N$ iteratively from Turyn's method, the proof of this construction follows if we apply \textit{Corollary} \ref{tur_th_pow} in every iteration.
	\end{IEEEproof}
	\vspace{0.1in}
	\renewcommand{\arraystretch}{0.6}
	\begin{table}
		\centering
		\tabcolsep=0.11cm
		\caption {$i_n$'s and $t_n$'s of the kernels of GCPs. \label{table_new45}}
		%
		\begin{tabular}{|c||c|}
			\hline
			Kernel case & $i_n$'s and $t_n$'s \\ \hline
			$K_{10}$ & $ \begin{aligned} & i_0=0,~ t_0=4, \\ & i_1=4,~ t_1=1, \\ & i_2=5,~ t_2=1,\\ & i_3=6,~ t_3=4 \end{aligned}$   \\ \hline
			$K_{26}$ & $\begin{aligned} & i_0=0,~ t_0=12, \\ & i_1=12,~ t_1=1,\\  & i_2=13,~ t_2=1,\\ & i_3=14,~ t_3=12 \end{aligned}$   \\ \hline
		\end{tabular}
	\end{table}

		\begin{example}\label{ex_101}
			Let $(\mathbf{a};\mathbf{b})$ be the GCP of length $100$ shown in (\ref{gcp100_1}), $x=1$ and $y=-1$. Consider $\mathbf{e}={\mathcal{I}(\mathbf{a},0,x)}$ and $\mathbf{f}={\mathcal{I}(\mathbf{b},0,y)}$, i.e.,
			\begin{equation}
			\left( \begin{matrix}
			\mathbf{e} \\
			\mathbf{f}
			\end{matrix} \right) = \left( X \Vert A \Vert B \Vert C \Vert D \right).
			\end{equation}
			where $A,B,C,D$ are given in (\ref{gcp100}) and $X=\left( \begin{smallmatrix}
			+ \\
			-
			\end{smallmatrix} \right)$.
			Then, $(\mathbf{e};\mathbf{f})$ is a length-$101$ Type-I OB-ZCP with a ZCZ width of $41$ because
			\begin{equation}
			\begin{array}{ccl}
			\Bigl ( \mid \rho_{\mathbf{e}}(\tau)+\rho_{\mathbf{f}}(\tau) \mid \Bigl )_{\tau=0}^{100} & = & (202,\mathbf{0}_{40},\mathbf{2}_{10},\mathbf{0}_{10},\mathbf{2}_{40}).
			\end{array}
			\end{equation}
		\end{example}
	\vspace{0.1in}
	\begin{example}\label{ex_10}
		Let $(\mathbf{a};\mathbf{b})$ be the GCP of length 100 shown in (\ref{gcp100_1}), $x=1$ and $y=1$. Then, $\mathbf{e}={\mathcal{I}(\mathbf{a},0,x)}$, $\mathbf{f}={\mathcal{I}(\mathbf{b},0,y)}$ can be expressed as follows.
		\begin{equation}
		\left( \begin{matrix}
		\mathbf{e} \\
		\mathbf{f}
		\end{matrix} \right) = \left( X \Vert A \Vert B \Vert C \Vert D \right),
		\end{equation}
		where $A,B,C,D$ are shown in (\ref{gcp100}) and $X=\left( \begin{smallmatrix}
		+ \\
		+
		\end{smallmatrix} \right)$.
		Here, $(\mathbf{e};\mathbf{f})$ is a length-$101$ Type-II OB-ZCP with a ZCZ width of $41$ because
		\begin{equation}
		\begin{array}{ccl}
		\Bigl ( \mid \rho_{\mathbf{e}}(\tau)+\rho_{\mathbf{f}}(\tau) \mid \Bigl )_{\tau=0}^{100} & = & (202,\mathbf{2}_{40},\mathbf{0}_{10},\mathbf{2}_{10},\mathbf{0}_{40}).
		\end{array}
		\end{equation}
	\end{example}
\begin{figure*}
	\centering
	\begin{minipage}[b]{.47\textwidth}
		\includegraphics[width=\textwidth]{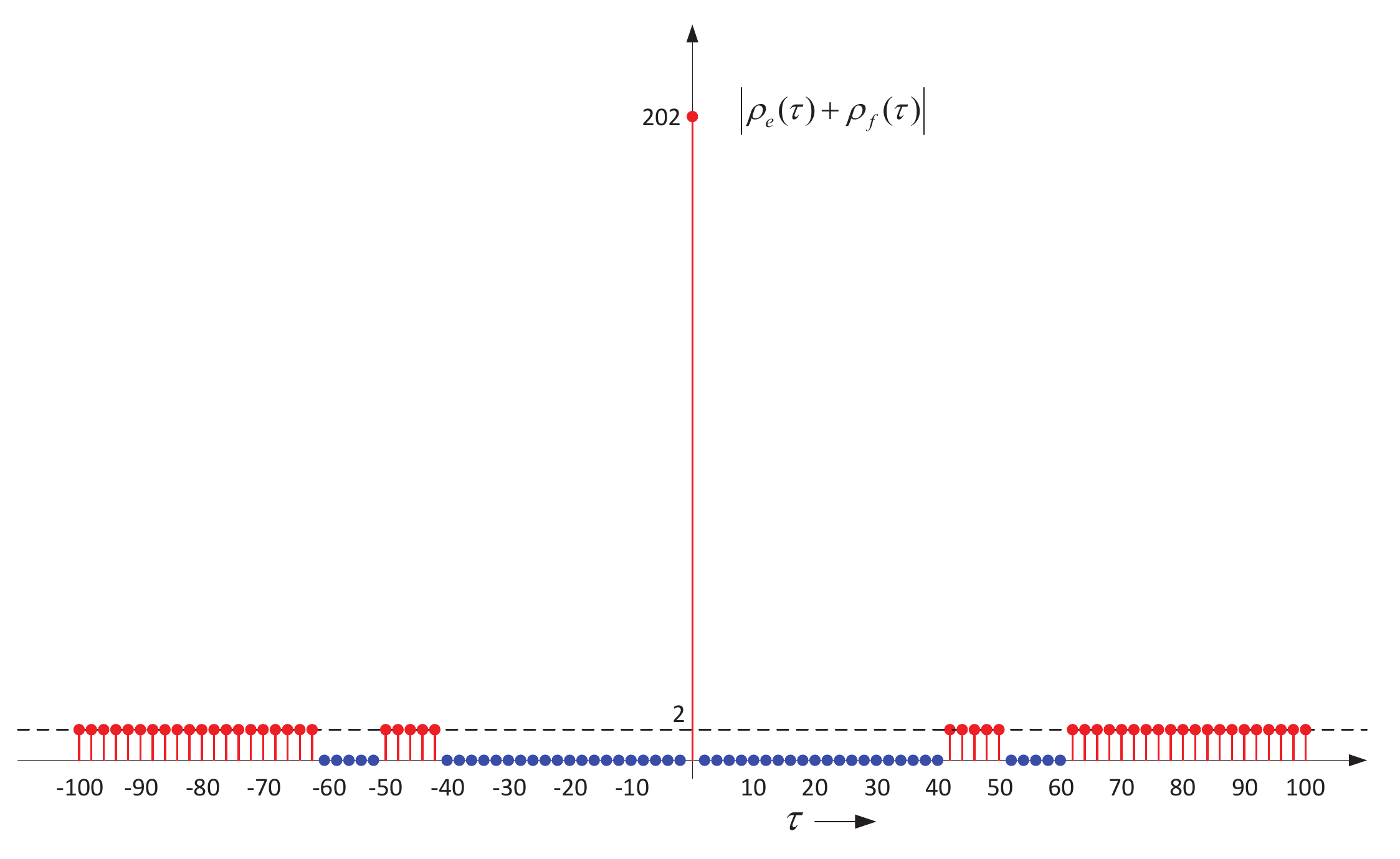}
		\caption{AACS magnitudes of OB-ZCP in \textit{Example \ref{ex_101}}.}\label{label-a}
	\end{minipage}\qquad
	\begin{minipage}[b]{.47\textwidth}
		\includegraphics[width=\textwidth]{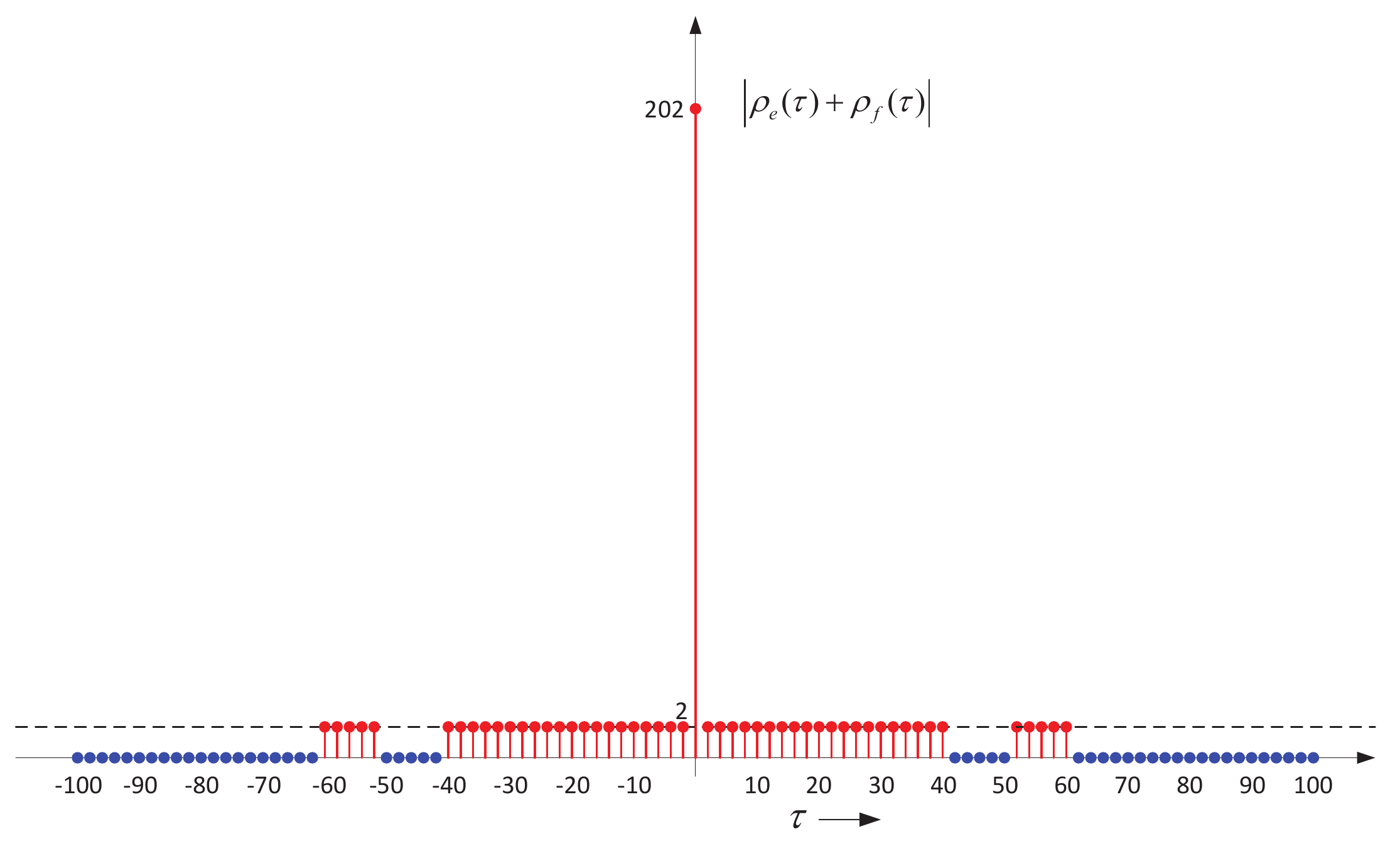}
		\caption{AACS magnitudes of OB-ZCP in \textit{Example \ref{ex_10}}.}\label{label-b}
	\end{minipage}
\end{figure*}
\vspace{0.1in}
	The following lemma gives the AACSs at different time-shifts when $x,y \in \mathbb{U}$ are inserted at the end position of the sequences $\mathbf{a}$ and $\mathbf{b}$, respectively.
	\vspace{0.1in}
	\begin{lemma}\label{th_end}
		Let ($\mathbf{a};\mathbf{b}$) be GCP of length $N$. If $\mathbf{e}={\mathcal{I}(\mathbf{a},N,x)}$ and $\mathbf{f}={\mathcal{I}(\mathbf{b},N,y)}$, where $x,y \in \mathbb{U}$, then we have
		\begin{equation}\label{mag_end}
		\rho_{\mathbf{e}}(\tau)+\rho_{\mathbf{f}}(\tau)=xa_{N-\tau}+yb_{N-\tau}, ~~~ 1\leq \tau <N.
		\end{equation}
	\end{lemma}
	\vspace{0.1in}
	\begin{IEEEproof}
		This proof can be easily obtained from (\ref{eqnend}).
	\end{IEEEproof}
	\vspace{0.1in}
	Note that (\ref{mag_end}) can be further reduced to
	\begin{equation}\label{th_end_eq}
	\rho_{\mathbf{e}}(\tau)+\rho_{\mathbf{f}}(\tau)=
	\begin{cases}
	0,~~ & \text{if}~xa_{N-\tau} \cdot yb_{N-\tau} = -1 \\
	\pm 2,~~ & \text{if}~xa_{N-\tau} \cdot yb_{N-\tau} = 1
	\end{cases}.
	\end{equation}
	\vspace{0.1in}
	Applying the \textit{Lemma} \ref{th_end}, \textit{Corollary} \ref{fac_2_tur}, and similar to \textit{Construction} \ref{th_pow_2_mag}, we have the following result.
	\vspace{0.1in}
	
	\begin{construction}\label{cor_pow_2}
		Let $(\mathbf{a};\mathbf{b})$ be a GCP [generated by (\ref{fac_2_tur_equ})] of length $N=2^\alpha M$, where $M=10^\beta 26^\gamma$, $\alpha,~\beta, \text{ and } \gamma$ are non-negative integers and $\alpha \geq 1$. If $x,y \in \mathbb{U}$ and $\mathbf{e}={\mathcal{I}(\mathbf{a},N,x)}$, $\mathbf{f}={\mathcal{I}(\mathbf{b},N,y)}$, then we have the following constructions:
		\begin{itemize}
		\item $(\mathbf{e};\mathbf{f})$ is a Type-I \textit{optimal} OB-ZCP with ZCZ width of $2^{\alpha -1} M +1$ satisfying the following,
		\begin{equation}
		\mid\rho_{\mathbf{e}}(\tau)+\rho_{\mathbf{f}}(\tau)\mid=
		\begin{cases}
		0,& \text{if }  0 < \tau \leq 2^{\alpha -1} M, \\
		2,& \text{if } 2^{\alpha -1} M < \tau <2^\alpha M.
		\end{cases}
		\end{equation}
		when $x$ and $y$ are of identical signs.
		\item $(\mathbf{e};\mathbf{f})$ is a Type-II \textit{optimal} OB-ZCP with ZCZ width of $2^{\alpha -1} M +1$ satisfying the following equation.
		\begin{equation}
		\mid\rho_{\mathbf{e}}(\tau)+\rho_{\mathbf{f}}(\tau)\mid=
		\begin{cases}
		2,& \text{if }  0 < \tau \leq 2^{\alpha -1} M, \\
		0,& \text{if } 2^{\alpha -1} M < \tau <2^\alpha M.
		\end{cases}
		\end{equation}
		when $x$ and $y$ are of different signs.
		\end{itemize}
	\end{construction}

	\vspace{0.1in}
	\begin{example}
		Let $(\mathbf{a};\mathbf{b})$ be the GCP of length 20 shown in (\ref{examp2_10}), $x=1 \text{ and } y=1$. Also, let $\mathbf{e}={\mathcal{I}(\mathbf{a},20,x)}$, $\mathbf{f}={\mathcal{I}(\mathbf{b},20,y)}$, i.e.,
		\begin{equation}
		\left( \begin{matrix}
		\mathbf{e} \\
		\mathbf{f}
		\end{matrix} \right) = \left( \begin{matrix}
		\textcolor{blue}{++-+-+--++}--+-----++ \textcolor{green}{+}\\
		\textcolor{blue}{++-+-+--++}++-+++++-- \textcolor{green}{+}
		\end{matrix} \right).
		\end{equation}
		Then, $(\mathbf{e},\mathbf{f})$ is a length-$21$ \textit{optimal} Type-I OB-ZCP with a ZCZ width of $11$ because
		\begin{equation}
		\begin{array}{ccl}
		\Bigl ( \mid \rho_{\mathbf{e}}(\tau)+\rho_{\mathbf{f}}(\tau) \mid \Bigl )_{\tau=0}^{20} & = & (42,\mathbf{0}_{10},\mathbf{2}_{10}).
		\end{array}
		\end{equation}
	\end{example}
	\vspace{0.1in}
	
		Recalling \textit{Corollary} \ref{tur_th_pow} and similar to \textit{Construction} \ref{th_pow_10_26}, we have the following construction.
		
		\vspace{0.1in}
		
		\begin{construction}\label{th_pow_10_266}
			Let $(\mathbf{a};\mathbf{b})$ be a GCP generated by \textit{Corollary} \ref{tur_th_pow} and of length $N^p$, where $N=10 \text{ or } 26$ and $p$ is a non-negative integer. If $x,y \in \mathbb{U}$ and $\mathbf{e}={\mathcal{I}(\mathbf{a},N,x)}$, $\mathbf{f}={\mathcal{I}(\mathbf{b},N,y)}$, then we have the following constructions:
			\begin{itemize}
				\item $(\mathbf{e};\mathbf{f})$ is a Type-I OB-ZCP with ZCZ width of $(i_0+t_0)N^{p-1} +1$ satisfying the following equation.
				\begin{equation}
				\begin{split}
				& \mid\rho_{\mathbf{e}}(\tau)+\rho_{\mathbf{f}}(\tau)\mid \\ &=
				\begin{cases}
				0,& \text{if }  i_0N^{p-1} < \tau \leq (i_0+t_0)N^{p-1}, \\
				2,& \text{if } i_1N^{p-1} < \tau \leq(i_1+t_1)N^{p-1} ,\\
				0,& \text{if } i_2N^{p-1} < \tau \leq (i_2+t_2)N^{p-1} , \\
				2,& \text{if } i_3N^{p-1} < \tau <(i_3+t_3)N^{p-1}=N^p,
				\end{cases}
				\end{split}
				\end{equation}
				when $x$ and $y$ are of identical signs.
				\item $(\mathbf{e};\mathbf{f})$ is a Type-II OB-ZCP with ZCZ width of $(i_0+t_0)N^{p-1} +1$, i.e.,
				\begin{equation}
				\begin{split}
				& \mid\rho_{\mathbf{e}}(\tau)+\rho_{\mathbf{f}}(\tau)\mid \\ & =
				\begin{cases}
				2,& \text{if }  i_0N^{p-1} < \tau \leq (i_0+t_0)N^{p-1}, \\
				0,& \text{if } i_1N^{p-1} < \tau \leq(i_1+t_1)N^{p-1} ,\\
				2,& \text{if } i_2N^{p-1} < \tau \leq (i_2+t_2)N^{p-1} , \\
				0,& \text{if } i_3N^{p-1} < \tau <(i_3+t_3)N^{p-1} =N^p,
				\end{cases}
				\end{split}
				\end{equation}
				when $x$ and $y$ are different signs.
			\end{itemize}
			Here $t_n$ is the number of consecutive columns of $K_N$, each having elements with identical/different sign (depending on the value of $t_n$), starting from the $i_n$-th column. The values of $(i_n,t_n)$ for $K_{10}$ and $K_{26}$ are given in Table \ref{table_new45}.
		\end{construction}
	
	\vspace{0.1in}
	
		 Based on \textit{Construction} \ref{th_pow_10_26} and \textit{Construction} \ref{th_pow_10_266} , we summarize more cases of Type-I OB-ZCPs in Table \ref{table_type1} and Type-II OB-ZCPs in Table \ref{table_type2}, with \textit{maximum} possible ZCZ widths which can be achieved by this method, when $N=10^\beta26^\gamma$.

	\begin{table*}
		\centering
		
		\caption{A Summary of Type-I OB-ZCPs With Large ZCZ Widths. \label{table_type1}}
		\resizebox{\textwidth}{!}{
			
			\begin{tabular}{|c||c|c|c|c|c||l|}
				\hline
				$\#$ & Sequence Length & $\mathbf{e}$ & $\mathbf{f}$ & Constraints & ZCZ width & $\mid\rho_{\mathbf{e}}(\tau)+\rho_{\mathbf{f}}(\tau)\mid$ \\ \hline \hline
				$1$ & $10^\beta $ & ${\mathcal{I}(\mathbf{a},0,x)}$ & ${\mathcal{I}(\mathbf{b},0,y)}$ & $x,y ~(\in \mathbb{U})$ are of different signs & $4 \times 10^{\beta -1} +1$ & $\begin{cases}
				0,& \text{if }  0 < \tau \leq 4 \times 10^{\beta -1}, \\
				2,& \text{if } 4 \times 10^{\beta -1} < \tau \leq 5 \times 10^{\beta -1} ,\\
				0,& \text{if } 5 \times 10^{\beta -1} < \tau \leq 6 \times 10^{\beta -1} , \\
				2,& \text{if } 6 \times 10^{\beta -1} < \tau < 10^\beta
				\end{cases}$
				\\  \hline
				$2$ & $26^\gamma $ & ${\mathcal{I}(\mathbf{a},0,x)}$ & ${\mathcal{I}(\mathbf{b},0,y)}$ & $x,y ~(\in \mathbb{U})$ are of different signs & $12 \times 26^{\gamma -1} +1$ & $\begin{cases}
				0,& \text{if }  0 < \tau \leq 12 \times 26^{\gamma -1}, \\
				2,& \text{if } 12 \times 26^{\gamma -1} < \tau \leq 13 \times 26^{\gamma -1} ,\\
				0,& \text{if } 13 \times 26^{\gamma -1} < \tau \leq 14 \times 26^{\gamma -1} , \\
				2,& \text{if } 14 \times 26^{\gamma -1} < \tau < 26^\gamma
				\end{cases}$
				\\  \hline
				$3$ & $10^\beta 26^\gamma $ & $\mathcal{I}(\mathbf{a},0,x)$ & $\mathcal{I}(\mathbf{b},0,y)$ & \shortstack{ $\beta \geq 1 \text{ and } \gamma \geq 1$, \\ $x,y ~(\in \mathbb{U})$ are of different signs} & $12 \times 26^{\gamma-1}10^\beta+1$ & $\begin{cases}
				0,& \text{if }  0 < \tau \leq 12 \times 26^{\gamma-1}10^\beta, \\
				2,& \text{if } 12 \times 26^{\gamma-1}10^\beta < \tau \leq 13 \times 26^{\gamma-1}10^\beta ,\\
				0,& \text{if } 13 \times 26^{\gamma-1}10^\beta < \tau \leq 14 \times 26^{\gamma-1}10^\beta , \\
				2,& \text{if } 14 \times 26^{\gamma-1}10^\beta < \tau < 26^\gamma 10^\beta
				\end{cases}$ \\ \hline
				
				$4$ & $10^\beta $ & ${\mathcal{I}(\mathbf{a},N,x)}$ & ${\mathcal{I}(\mathbf{b},N,y)}$ & $x,y ~(\in \mathbb{U})$ are of identical sign & $4 \times 10^{\beta -1} +1$ & $\begin{cases}
				0,& \text{if }  0 < \tau \leq 4 \times 10^{\beta -1}, \\
				2,& \text{if } 4 \times 10^{\beta -1} < \tau \leq 5 \times 10^{\beta -1} ,\\
				0,& \text{if } 5 \times 10^{\beta -1} < \tau \leq 6 \times 10^{\beta -1} , \\
				2,& \text{if } 6 \times 10^{\beta -1} < \tau < 10^\beta
				\end{cases}$
				\\  \hline
				$5$ & $26^\gamma$ & ${\mathcal{I}(\mathbf{a},N,x)}$ & ${\mathcal{I}(\mathbf{b},N,y)}$ & $x,y ~(\in \mathbb{U})$ are of identical sign & $12 \times 26^{\gamma -1} +1$ & $\begin{cases}
				0,& \text{if }  0 < \tau \leq 12 \times 26^{\gamma -1}, \\
				2,& \text{if } 12 \times 26^{\gamma -1} < \tau \leq 13 \times 26^{\gamma -1} ,\\
				0,& \text{if } 13 \times 26^{\gamma -1} < \tau \leq 14 \times 26^{\gamma -1} , \\
				2,& \text{if } 14 \times 26^{\gamma -1} < \tau < 26^\gamma
				\end{cases}$
				\\  \hline
				$6$ & $10^\beta 26^\gamma $ & $\mathcal{I}(\mathbf{a},N,x)$ & $\mathcal{I}(\mathbf{b},N,y)$ & \shortstack{ $\beta \geq 1 \text{ and } \gamma \geq 1$, \\ $x,y (\in \mathbb{U})$ are of identical sign} & $12 \times 26^{\gamma-1}10^\beta+1$ & $\begin{cases}
				0,& \text{if }  0 < \tau \leq 12 \times 26^{\gamma-1}10^\beta, \\
				2,& \text{if } 12 \times 26^{\gamma-1}10^\beta < \tau \leq 13 \times 26^{\gamma-1}10^\beta ,\\
				0,& \text{if } 13 \times 26^{\gamma-1}10^\beta < \tau \leq 14 \times 26^{\gamma-1}10^\beta , \\
				2,& \text{if } 14 \times 26^{\gamma-1}10^\beta < \tau < 26^\gamma 10^\beta
				\end{cases}$ \\ \hline
				\hline
			\end{tabular}
		}
	\end{table*}

	\vspace{0.1in}

	\begin{table*}
		\centering
		
		\caption{A Summary of Type-II OB-ZCPs With Large ZCZ Widths. \label{table_type2}}
		\resizebox{\textwidth}{!}{
			
			\begin{tabular}{|c||c|c|c|c|c||l|}
				\hline
				$\#$ & Sequence Length & $\mathbf{e}$ & $\mathbf{f}$ & Constraints & ZCZ width & $\mid\rho_{\mathbf{e}}(\tau)+\rho_{\mathbf{f}}(\tau)\mid$ \\ \hline \hline
				$1$ & $10^\beta $ & ${\mathcal{I}(\mathbf{a},0,x)}$ & ${\mathcal{I}(\mathbf{b},0,y)}$ & $x,y ~(\in \mathbb{U})$ are of identical signs & $4 \times 10^{\beta -1} +1$ & $\begin{cases}
				2,& \text{if }  0 < \tau \leq 4 \times 10^{\beta -1}, \\
				0,& \text{if } 4 \times 10^{\beta -1} < \tau \leq 5 \times 10^{\beta -1} ,\\
				2,& \text{if } 5 \times 10^{\beta -1} < \tau \leq 6 \times 10^{\beta -1} , \\
				0,& \text{if } 6 \times 10^{\beta -1} < \tau < 10^\beta
				\end{cases}$
				\\  \hline
				$2$ & $26^\gamma $ & ${\mathcal{I}(\mathbf{a},0,x)}$ & ${\mathcal{I}(\mathbf{b},0,y)}$ & $x,y ~(\in \mathbb{U})$ are of identical signs & $12 \times 26^{\gamma -1} +1$ & $\begin{cases}
				2,& \text{if }  0 < \tau \leq 12 \times 26^{\gamma -1}, \\
				0,& \text{if } 12 \times 26^{\gamma -1} < \tau \leq 13 \times 26^{\gamma -1} ,\\
				2,& \text{if } 13 \times 26^{\gamma -1} < \tau \leq 14 \times 26^{\gamma -1} , \\
				0,& \text{if } 14 \times 26^{\gamma -1} < \tau < 26^\gamma
				\end{cases}$
				\\  \hline
				$3$ & $10^\beta 26^\gamma $ & $\mathcal{I}(\mathbf{a},0,x)$ & $\mathcal{I}(\mathbf{b},0,y)$ & \shortstack{ $\beta \geq 1 \text{ and } \gamma \geq 1$, \\ $x,y ~(\in \mathbb{U})$ are of identical signs} & $12 \times 26^{\gamma-1}10^\beta+1$ & $\begin{cases}
				2,& \text{if }  0 < \tau \leq 12 \times 26^{\gamma-1}10^\beta, \\
				0,& \text{if } 12 \times 26^{\gamma-1}10^\beta < \tau \leq 13 \times 26^{\gamma-1}10^\beta ,\\
				2,& \text{if } 13 \times 26^{\gamma-1}10^\beta < \tau \leq 14 \times 26^{\gamma-1}10^\beta , \\
				0,& \text{if } 14 \times 26^{\gamma-1}10^\beta < \tau < 26^\gamma 10^\beta
				\end{cases}$ \\ \hline
				
				$4$ & $10^\beta $ & ${\mathcal{I}(\mathbf{a},N,x)}$ & ${\mathcal{I}(\mathbf{b},N,y)}$ & $x,y ~(\in \mathbb{U})$ are of different signs & $4 \times 10^{\beta -1} +1$ & $\begin{cases}
				2,& \text{if }  0 < \tau \leq 4 \times 10^{\beta -1}, \\
				0,& \text{if } 4 \times 10^{\beta -1} < \tau \leq 5 \times 10^{\beta -1} ,\\
				2,& \text{if } 5 \times 10^{\beta -1} < \tau \leq 6 \times 10^{\beta -1} , \\
				0,& \text{if } 6 \times 10^{\beta -1} < \tau < 10^\beta
				\end{cases}$
				\\  \hline
				$5$ & $26^\gamma $ & ${\mathcal{I}(\mathbf{a},N,x)}$ & ${\mathcal{I}(\mathbf{b},N,y)}$ & $x,y ~(\in \mathbb{U})$ are of different signs & $12 \times 26^{\gamma -1} +1$ & $\begin{cases}
				2,& \text{if }  0 < \tau \leq 12 \times 26^{\gamma -1}, \\
				0,& \text{if } 12 \times 26^{\gamma -1} < \tau \leq 13 \times 26^{\gamma -1} ,\\
				2,& \text{if } 13 \times 26^{\gamma -1} < \tau \leq 14 \times 26^{\gamma -1} , \\
				0,& \text{if } 14 \times 26^{\gamma -1} < \tau < 26^\gamma
				\end{cases}$
				\\  \hline
				$6$ & $10^\beta 26^\gamma $ & $\mathcal{I}(\mathbf{a},N,x)$ & $\mathcal{I}(\mathbf{b},N,y)$ & \shortstack{ $\beta \geq 1 \text{ and } \gamma \geq 1$, \\ $x,y ~(\in \mathbb{U})$ are of different signs} & $12 \times 26^{\gamma-1}10^\beta+1$ & $\begin{cases}
				2,& \text{if }  0 < \tau \leq 12 \times 26^{\gamma-1}10^\beta, \\
				0,& \text{if } 12 \times 26^{\gamma-1}10^\beta < \tau \leq 13 \times 26^{\gamma-1}10^\beta ,\\
				2,& \text{if } 13 \times 26^{\gamma-1}10^\beta < \tau \leq 14 \times 26^{\gamma-1}10^\beta , \\
				0,& \text{if } 14 \times 26^{\gamma-1}10^\beta < \tau < 26^\gamma 10^\beta
				\end{cases}$ \\ \hline

				\hline
			\end{tabular}
		}
	\end{table*}

	\vspace{0.1in}
	
	Next, we will investigate the AACSs when $x,y \in \mathbb{U}$ are respectively inserted at the middle position of sequences $\mathbf{a}$ and $\mathbf{b}$. We need the following lemma.
		\vspace{0.1in}
		\begin{lemma}\label{general_middle_insertion}
			Let $(\mathbf{a};\mathbf{b})$ be a GCP of length $N$ and $(\mathbf{c};\mathbf{d})$ be a GCP of length $2N$, constructed according to \textit{corollary} \ref{fac_2_tur}. Let $\mathbf{c}^1$ and $\mathbf{c}^2$ denotes the first- and second- halves of $\mathbf{c}$ respectively, as in (\ref{seq_rep}). Similarly, let us define $\mathbf{d}^1$ and $\mathbf{d}^2$ for $\mathbf{d}$. Then,
			\begin{equation}\label{lemma_middle_1}
			\rho_{\mathbf{c}^1}(\tau)+\rho_{\mathbf{c}^2}(\tau)+\rho_{\mathbf{d}^1}(\tau)+\rho_{\mathbf{d}^2}(\tau)=0, ~~ \text{for} ~\tau \neq 0.
			\end{equation}
			In addition,
			\begin{equation}\label{lemma_middle_2}
			\rho_{\mathbf{c}^2,\mathbf{c}^1}(\tau)+\rho_{\mathbf{d}^2,\mathbf{d}^1}(\tau)=0.
			\end{equation}
		\end{lemma}
		\vspace{0.1in}
		\begin{IEEEproof}
			The proof is similar to that of [\ref{zilong_obzcp}, \textit{Lemma 6}] and thus omitted here.
		\end{IEEEproof}
		\vspace{0.1in}
		
		Using \textit{Lemma} \ref{general_middle_insertion} and \textit{Corollary} \ref{fac_2_tur}, we have the following construction.
		
		\vspace{0.1in}
		
		\begin{construction}
			Let $(\mathbf{c};\mathbf{d})$ be a GCP of length $2N$, constructed according to \textit{Lemma \ref{general_middle_insertion}}, where $N=2^\alpha10^\beta26^\gamma$, and $\alpha,\beta,\gamma$ are non-negative integers. If $x,y \in \mathbb{U}$ and $\mathbf{e}=\mathcal{I}(\mathbf{c},N,x)$ and $\mathbf{f}=\mathcal{I}(\mathbf{d},N,y)$, then $(\mathbf{e};\mathbf{f})$ is an \textit{optimal} Type-II OB-ZCP of length $2N+1$.
		\end{construction}
		\vspace{0.1in}
		\begin{IEEEproof}
			From (\ref{eqnmiddle}) we have the following cases:
			\begin{itemize}
				\item case I: $0 < \tau < N$:
				\begin{equation}
				\begin{split}
				&\rho_{\mathbf{e}}(\tau)+\rho_{\mathbf{f}}(\tau) = \\
				&\qquad \rho_{\mathbf{c}^1}(\tau)+xc_{N-\tau}+\rho_{\mathbf{c}^2,\mathbf{c}^1}(N-\tau+1)+\\
				& \qquad xc_{N+\tau-1} +\rho_{\mathbf{c}^2}(\tau)+ \rho_{\mathbf{d}^1}(\tau)+yd_{N-\tau} +   \\
				& \qquad  \rho_{\mathbf{d}^2,\mathbf{d}^1}(N-\tau+1)+ yd_{N+\tau-1}+  \rho_{\mathbf{d}^2}(\tau) \\
				& = xc_{N-\tau}+xc_{N+\tau-1} +yd_{N-\tau} +yd_{N+\tau-1} \\
				& \qquad \qquad \qquad \qquad \qquad \quad \qquad \qquad (\text{using \textit{Lemma} \ref{general_middle_insertion}}) \\
				& = \pm 2 \quad \qquad \qquad \quad (\text{using \textit{Lemma} \ref{prop_quad}, since } x,y \in \mathbb{U}).
				\end{split}
				\end{equation}
				
				\item case II: $\tau = N$: \\
				\begin{equation*}
				\begin{split}
				&  \rho_{\mathbf{e}}(\tau)+\rho_{\mathbf{f}}(\tau)= \\
				& \qquad xc_{N-\tau}+\rho_{\mathbf{c}^2,\mathbf{c}^1}(N-\tau+1)+xc_{N+\tau-1}  \\
				& \qquad + yd_{N-\tau}+\rho_{\mathbf{d}^2,\mathbf{d}^1}(N-\tau+1)+yd_{N+\tau-1} \\
				& \qquad \qquad \qquad \qquad \qquad \quad \qquad \qquad (\text{using \textit{Lemma} \ref{general_middle_insertion}})
				\end{split}
				\end{equation*}
				\begin{equation}
				\begin{split}
				& = xc_{N-\tau}+xc_{N+\tau-1}+yd_{N-\tau}+yd_{N+\tau-1} \\
				& = \pm 2 \quad \qquad \qquad \quad (\text{using \textit{Lemma} \ref{prop_quad}, since } x,y \in \mathbb{U}).
				\end{split}
				\end{equation}
				
				\item case III: $N < \tau < 2N+1$: \\
				\begin{equation}
				\begin{split}
				& \rho_{\mathbf{e}}(\tau)+\rho_{\mathbf{f}}(\tau)  = \\
				& \qquad \rho_{\mathbf{c}^2,\mathbf{c}^1}(N-\tau+1)
				+ \rho_{\mathbf{d}^2,\mathbf{d}^1}(N-\tau+1) =0 \\
				& \qquad \qquad \qquad \qquad \qquad \quad \qquad \qquad  (\text{using \textit{Lemma} \ref{general_middle_insertion}}).
				\end{split}
				\end{equation}
			\end{itemize}
			This shows that $(\mathbf{e};\mathbf{f})$ has ZCZ width of $N+1$ and the AACS magnitude at every out-of-zone time-shift is $2$. Hence, it is \textit{optimal}.
		\end{IEEEproof}
		\vspace{0.1in}
		\begin{example}\label{ex_2_10_mid}
			Let $(\mathbf{c};\mathbf{d})$ be the GCP of length 20 shown in (\ref{examp2_10}), $x=1$ and $y=1$, Then, $\mathbf{e}=\mathcal{I}(\mathbf{c},10,x)$, and $\mathbf{f}=\mathcal{I}(\mathbf{d},10,y)$ are
		{\color{black}	\begin{equation}
			\left( \begin{matrix}
			\mathbf{e} \\
			\mathbf{f}
			\end{matrix} \right) = \left( \begin{matrix}
			\textcolor{blue}{++-+-+--++}\textcolor{green}{+}--+-----++\\
			\textcolor{blue}{++-+-+--++}\textcolor{green}{+}++-+++++--
			\end{matrix} \right).
			\end{equation} }
			Here, $(\mathbf{e};\mathbf{f})$ is an \textit{optimal} Type-II OB-ZCP of length 21 with a ZCZ width of $11$ because
			\begin{equation}
			\begin{array}{ccl}
			\Bigl ( \mid \rho_{\mathbf{e}}(\tau)+\rho_{\mathbf{f}}(\tau) \mid \Bigl )_{\tau=0}^{20} & = & (42,\mathbf{2}_{10},\mathbf{0}_{10}).
			\end{array}
			\end{equation}
		\end{example}
		\vspace{0.1in}
		\begin{remark}
			During computer search, we got \textit{optimal} Type-II OB-ZCPs of length $11$ when insertion method is employed on $K_{10}$ (see \textit{Example \ref{ex_10_computersearch}}). But we did not get \textit{optimal} OB-ZCPs for cases when the GCP is of length $10^\beta, ~ \beta > 1$. Also, we did not find \textit{optimal} OB-ZCP, when the GCP length of the form $26^\gamma,~ \text{and } 10^\beta26^\gamma$.
		\end{remark}
		\vspace{0.1in}
		\begin{example}\label{ex_10_computersearch}
			Let $(\mathbf{c};\mathbf{d})$ be $K_{10}$, $x=1$ and $y=1$, Then, $\mathbf{e}=\mathcal{I}(\mathbf{c},5,x)$, and $\mathbf{f}=\mathcal{I}(\mathbf{d},4,y)$ are
		{\color{black}	\begin{equation}
			\left( \begin{matrix}
			\mathbf{e} \\
			\mathbf{f}
			\end{matrix} \right) = \left( \begin{matrix}
			{\color{blue}++-+-}{\color{green}+}+--++\\
			{\color{blue}++-+}{\color{green}+}++++--
			\end{matrix} \right).
			\end{equation} }
			Here, $(\mathbf{e};\mathbf{f})$ is an \textit{optimal} Type-II OB-ZCP of length 11 with a ZCZ width of $6$ because
			\begin{equation}
			\begin{array}{ccl}
			\Bigl ( \mid \rho_{\mathbf{e}}(\tau)+\rho_{\mathbf{f}}(\tau) \mid \Bigl )_{\tau=0}^{10} & = & (22,\mathbf{2}_{5},\mathbf{0}_{5}).
			\end{array}
			\end{equation}
	\end{example}

	\section{CONCLUSION}
	
	In this paper, we have explored various \textit{intrinsic} properties of binary GCPs which are constructed from Turyn's method.
	Specifically, by exploring Turyn's method to construct binary GCPs of lengths $2^\alpha10^\beta26^\gamma$, we are able to identify which column of GCP has identical sign (and which has opposite) when it is arranged as a two-dimensional matrix containing two row sequences. These properties allow us to construct \textit{optimal} OB-ZCPs (Type-I and Type-II) of lengths $2^\alpha 10^\beta 26^\gamma +1$ (with $\alpha \geq 1$) by proper insertion of GCPs. For OB-ZCPs of lengths $10^\beta+1,~ 26^\gamma+1, \text{and } 10^\beta 26^\gamma+1$, we have shown that the \textit{largest} possible ZCZ widths are $4 \times 10^{\beta-1} +1$, $12 \times 26^{\gamma -1}+1$ and $12 \times 10^\beta 26^{\gamma -1}+1$, also by taking advantage of these \textit{intrinsic} structure properties. An interesting future work of this research is to find some systematic constructions of \textit{optimal} OB-ZCPs having lengths not of the form $2^{\alpha}10^{\beta}26^{\gamma}+1$.



\end{document}